\documentclass[12pt,preprint,times]{aastex}

\newcommand{\pref}{\protect\ref}
\newcommand{\brr}{$\langle B_{||} \rangle_{r=20}$}

\begin{document}

\shorttitle{Coronal Dimming Evolution}
\shortauthors{S.~W. McIntosh et~al.}
\title{The Post-Eruptive Evolution of a Coronal Dimming}
\author{Scott W. McIntosh\altaffilmark{1,2,4}, Robert J. Leamon\altaffilmark{3}, Alisdair R. Davey\altaffilmark{1}, Meredith J. Wills-Davey\altaffilmark{1}}
\altaffiltext{1}{Southwest Research Institute, Department of Space Studies, 1050 Walnut St, Suite 400, Boulder, CO 80302 USA}
\altaffiltext{2}{Visitor at the High Altitude Observatory, National Center for Atmospheric Research, P.O. Box 3000, Boulder, CO 80307 USA}
\altaffiltext{3}{Adnet Systems Inc., NASA Goddard Space Flight Center, Code 671.1, Greenbelt, MD 20771 USA}
\altaffiltext{4}{Corresponding Author: mcintosh@boulder.swri.edu}

\begin{abstract}
We discuss the post-eruptive evolution of a ``coronal dimming'' based on observations of the EUV corona from the Solar and Heliospheric Observatory and the Transition Region and Coronal Explorer. This discussion highlights the roles played by magnetoconvection-driven magnetic reconnection and the global magnetic environment of the plasma in the ``filling'' and apparent motion of the region following the eruption of a coronal mass ejection (CME). A crucial element in our understanding of the dimming region evolution is developed by monitoring the disappearance and reappearance of bright TRACE ``moss'' around the active region giving rise to the CME. We interpret the change in the TRACE moss as a proxy of the changing coronal magnetic field topology behind the CME front. We infer that the change in global magnetic topology also results in a shift of energy balance in the process responsible for the production of the moss emission while the coronal magnetic topology evolves from closed, to open and back to closed again because, following the eruption, the moss reforms around the active region in almost exactly its pre-event configuration. As a result of the moss evolution, combining our discussion with recent spectroscopic results of an equatorial coronal hole, we suggest that the interchangeable use of the term ``transient coronal hole'' to describe a coronal dimming is more than just a simple coincidence.
\end{abstract}


\section{Introduction}
Coronal dimmings, or ``transient coronal holes'' as they are sometimes known, are a source of great interest in the solar physics community \cite[e.g.,][]{Rust1983, Kahler2001, Attrill2006}. Following their initial observation in the X-Ray corona with Skylab \cite[][]{Rust1983}, interest in their behavior has gained some momentum and many studies have been compiled illustrating a link between the occurrence of coronal dimmings and coronal mass ejections \cite[CMEs; see, e.g.,][]{Forbes2000}. Indeed, some 29\% of a recent survey of halo CMEs were clearly associated with coronal dimmings (A. Reinard \-- private communication), implying that roughly twice that percentage were related to front-side CMEs \-- far higher than the consistently observed 40\% CME to flare correlation \cite[][]{Munro1979, Webb1987, Harrison1995}. Following the CME eruption the large-scale transient dimming features are observed as a rapid evacuation of coronal material by EUV imaging instruments; the Transition Region and Coronal Explorer \cite[TRACE;][]{Handy1999} and the Extreme-ultraviolet Imaging Telescope \cite[EIT;][]{Boudine1995} on the Solar and Heliospheric Observatory \cite[SOHO;][]{Fleck1995}.

In this Paper we discuss the post-CME evolution of one coronal dimming (observed 2006 July 6) and speculate that most, if not all, other coronal dimmings will obey the same characteristic behavior \cite[including, for example, the well-studied dimming event of 1997 May 12, e.g.,][]{Webb2000}. To aid in the analysis we draw upon results published recently \cite[][]{McIntosh2006a, McIntosh2006b, Jefferies2006, McIntosh2007a} that have demonstrated how the magnetic topology of the plasma influences magnetoconvection-driven magnetic reconnection making it into a relentless, substantial, ``basal'' energy source for the solar atmosphere. The basic spatial unit of energy release is a thin magnetic feature resulting from the destruction of emerging and advecting magnetic flux that provides the mass, thermal and kinetic energies to the plasma in the quiet Sun, coronal holes and active regions. It appears that the rate and amount of energy (and mass) that is released into each plasma environment is controlled by the interaction of the emerging flux and the magnetic field on scales typical of supergranules ($\sim$20Mm). How these small-scale (``spicular''; see below)  mass and energy release events are driven by the relentless magneto-convection is controlled by the supergranular flux balance while the global topology of the coronal plasma has a large influence on how the energy is distributed; in the zero-mean field ``closed'' quiet-Sun the bulk of the energy released by reconnection is used thermally while, in the unbalanced mean-field ``open'' coronal hole is it delivered kinetically to the plasma. 

The result presented below suggests that calling the observed dimming phenomena a ``transient coronal hole'' is more than just a ``throw-away'' designation. The magnetic and energetic properties of the plasma perform in exactly same way as a static long-lived coronal hole except that is decays considerably faster. Further, we believe that the evolution of the dimming region and the nearly instantaneous release of a fast wind stream must have a significant impact on the initiating CME. With the top of the ``atmospheric pressure cooker'' removed by the eruption of the CME the energy of the ``boiling'' plasma below is released (in the form of mass loading and kinetic energy) immediately behind the front of the CME \cite[Rust 1983 and observed in the extreme by][]{Skoug2004} until large scale reconnection ``severs'' the CME from the Sun and the coronal magnetic field begins to close again \cite[][]{Forbes2000, Lin2000}.

\section{Data}
We use EUV observations of the corona in the 171\AA{} passband of TRACE and the 195\AA{} passband of SOHO/EIT that reflect the emitting state of the coronal plasma at $\sim$1MK \cite[][]{Handy1999} and $\sim$1.5MK \cite[][]{Boudine1995} mean temperatures respectively. On the day of the coronal dimming, 2006 July 6, SOHO/EIT was operating in its full-disk, 12-minute cadence, ``CME Watch'' observing mode and TRACE was observing NOAA AR10898 (source of the CME and dimming) for the full day operating with a 80s mean cadence and a 512\arcsec x512\arcsec{} field of view. We use data that cover the full duration of the CME and dimming in EIT (2006/07/06 07:13UT - 2006/07/07 12:48UT) and more limited coverage with TRACE (08:00 - 19:54UT on 2006/07/06) following the filament eruption that signals the start of the CME at $\sim$07:48UT. All of the data presented pictorially (and as movies) are co-aligned and de-rotated to the same reference time (07:13UT- the start of the EIT sequence) using Dominic Zarro's IDL mapping software ({\url http://orpheus.nascom.nasa.gov/$\sim$zarro/idl/maps.html}).

In addition to the EUV imaging data we must use the full-disk line-of-sight data from the Michelson Doppler Imager (MDI; Scherrer et al. 1995) that is acquired at regular 96-minute intervals over the course of the event (2006/07/06 07:59UT - 2006/07/07 07:59UT). While not using the magnetograms directly in our analysis we will employ two derivatives, the 20Mm supergranule averaged magnetogram (\brr) and the ``Magnetic Range of Influence'' (MRoI) that were presented by \cite{McIntosh2006a} and developed further in the analysis of \cite{McIntosh2006b, McIntosh2007a}. The \brr{} magnetic diagnostic represents the mean magnetic flux over a spatial scale of 20Mm (assumed to be commensurate with that of a typical supergranule) while the MRoI determines the distance needed from a particular flux concentration to encompass enough opposite polarity flux to consider the net magnetic flux to be in balance and can be loosely interpreted as the degree of topological ``closedness'' (small value of MRoI) or ``openness'' (large value of MRoI). The magnetic diagnostics are also de-rotated and co-aligned with the EIT 195\AA{} data at 07:13UT on 2006 July 6. 

\section{Analysis, Discussion and Interpretation}
\begin{figure}
\plotone{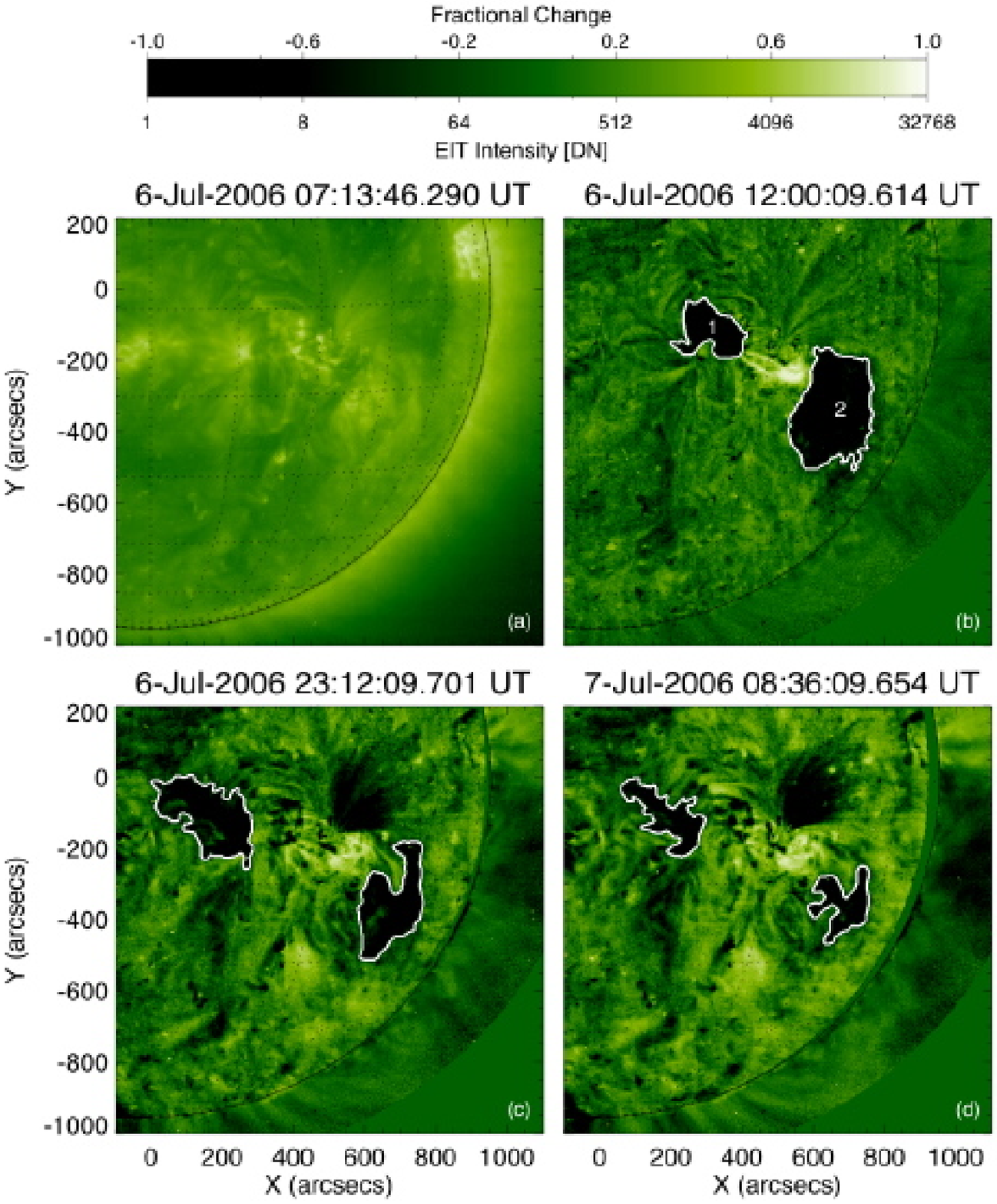}
\caption{SOHO/EIT 195\AA{} evolution of the coronal dimming. Panels B through D show the percentage difference in the EIT emission (from the baseline image presented in panel A) over the course of the event. The contours shown in these panels represent a 40\% reduction in intensity from the base image. [See the electronic edition of the Journal for an mpeg animation of this figure, where panels A through D are the 1st, 25th, 76th and 121st frames.] \label{fig1}}
\end{figure}

In Figure~\pref{fig1} we show four snapshots in the evolution of the dimming taken by EIT that are frames in the associated online movie. From left to right (top to bottom) the panels of the figure show the ``Before'' Intensity baseline image (07:13UT; panel A), the percentage difference (computed from the image in panel A) images at the ``Maximum Percentage Dimming'' (12:00UT; panel B), ``Maximum Area/Extent'' (23:12UT; panel C) and 24 hours post eruption (08:36UT; panel D) phases of the dimming evolution. In panels B, C and D we show the contour corresponding to a 40\% reduction in intensity from the (pre-event) baseline image. For convenience and clarity in the following discussion we have labeled the North-East and South-West dimming ``lobes'' 1 and 2 respectively. In Fig.~\pref{fig1}, the dimming regions appear to slowly migrate outward from the active region with lobe 2 stalling rapidly to the Southwest of the active region (by 09:19UT) while lobe 1 continues to evolve and move slowly in a North Easterly direction until about 17:00UT.

\begin{figure}
\plotone{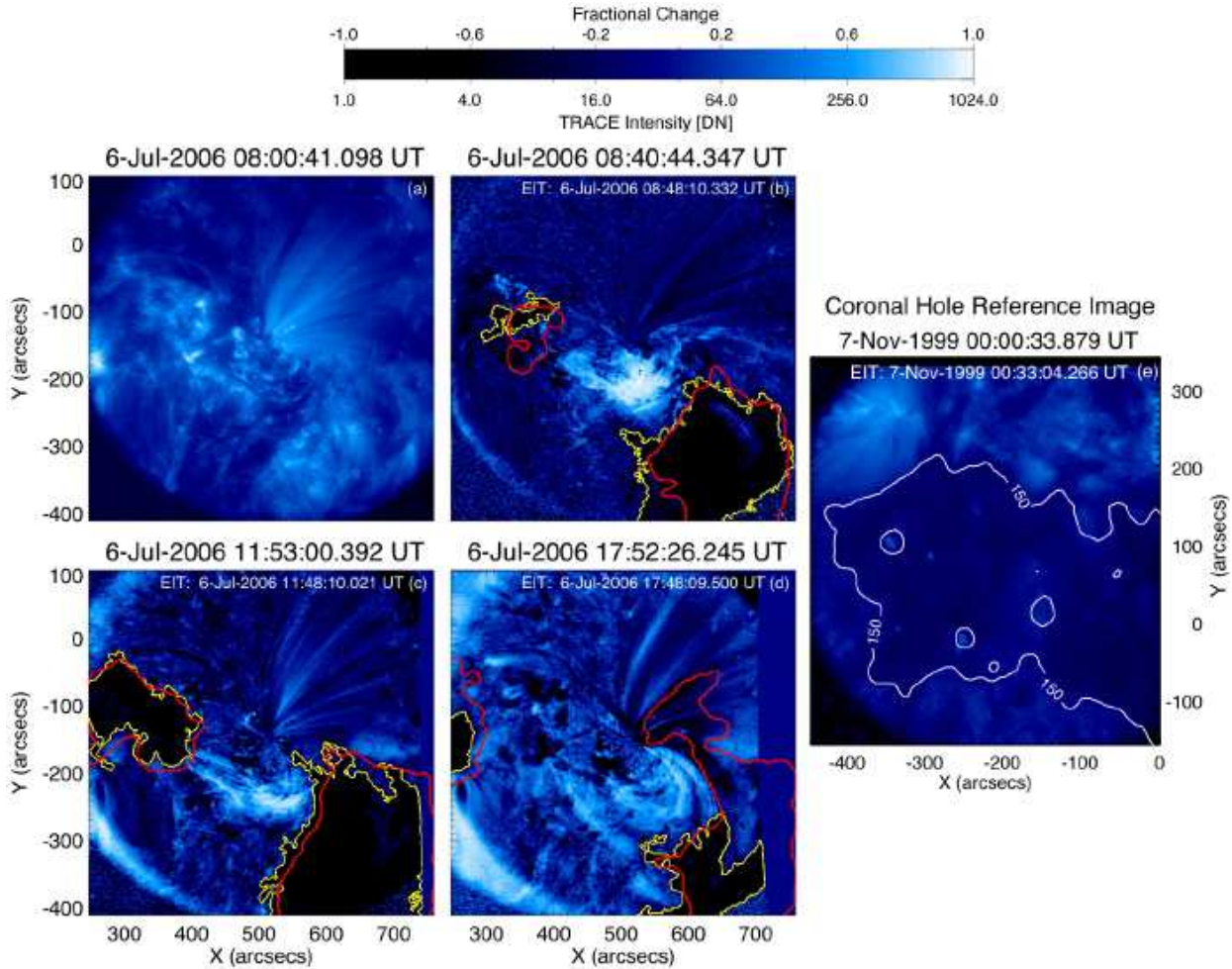}
\caption{TRACE 171\AA{} evolution of the coronal dimming. Panels B through D show the percentage difference in the TRACE emission (from the baseline image presented in panel A) over the course of the event. The contours shown in these panels represent a 40\% reduction in the image (yellow) and the corresponding 40\% reduction in the EIT emission closest to the time of the frame (red) that should be compared with those in Fig. 1. Panel E shows the TRACE 171\AA{} emission from an equatorial coronal hole (McIntosh et al. 2006), the white contour is the coronal hole boundary determined from the EIT 195\AA{} emission. The times of the EIT frames used are shown in the upper right of each panel. [See the electronic edition of the Journal for an mpeg animation of this figure.] \label{fig2}}
\end{figure}

In Figure~\pref{fig2}, in the slightly cooler coronal plasma that is observed at 171\AA{} with TRACE (and at significantly higher spatial resolution), the dimming begins at 08:18UT where the opposing poles of the active region show the rapid brightening above plage regions almost simultaneously [400\arcsec, -100\arcsec] and [620\arcsec, -300\arcsec] which, as time progresses and the filament finally clears the field of view (08:40UT, panel B) we begin to see the significant disappearance of ``moss'' \cite{Berger1999} from the same regions\footnote{The bright TRACE moss emission in 171\AA{} is formed at the base of the hot, high-pressure, coronal loops that are anchored in plage surrounding and closed overlying magnetic fields of active regions \cite[][]{Fletcher1999} and its appearance has been linked to the mass deposition by chromospheric spicules and the heating of the solar corona \cite[][]{DePontieu1999, DePontieu2006}.}. The leading edge of the moss ``erosion'' appears as an enhancement (as much as 40\%) in the percentage difference image, and is visible centered at [325\arcsec, -75\arcsec] in panel B. As the moss is gradually removed from the region and the image dims we only see enhanced emission in the heart of the active region (post-eruption loops) and across the neutral line threading it (10:43UT). The removal of the moss at each pole of the active region can be easily interpreted as the global coronal magnetic field ``opening'' as it is tied to the CME front \cite[e.g.,][]{Forbes2000, Lin2000}. At this phase of the dimming the 171\AA{} emission in the lobes resembles that of a coronal hole - void of moss \cite[cf. panel E, the TRACE 171\AA{} emission from the equatorial coronal hole studied by][]{McIntosh2006b}. 

After 10:43UT we see that the moss begins to rapidly develop again in dimming lobe 1 spreading from West to East (right to left) in outward from the active region and to the South of the neutral line reaching the pre-eruption configuration at $\sim$18:43UT. Just as we have interpreted the removal of moss as the opening of the magnetic field we can similarly interpret the reintroduction of moss to the system as the post-CME reconnection and closure of the global coronal field above the active region. This results from a change in how the mass and heat are distributed in the 1.5MK corona \cite[][]{McIntosh2006a, McIntosh2006b,McIntosh2007a}. At the same time we note that dimming lobe 2 does not fill as vigorously with moss, remaining almost in the immediate post-CME state. We have noted this behavior in the EIT data above. While lobe 2 stalls in place and appears to very slowly fill, lobe 1 appears to fill rapidly from the Eastern side of the active region. Again we note that the filling from the active region side gives the appearance that dimming lobe 1 migrates or moves slowly to the North East.

\begin{figure}
\plotone{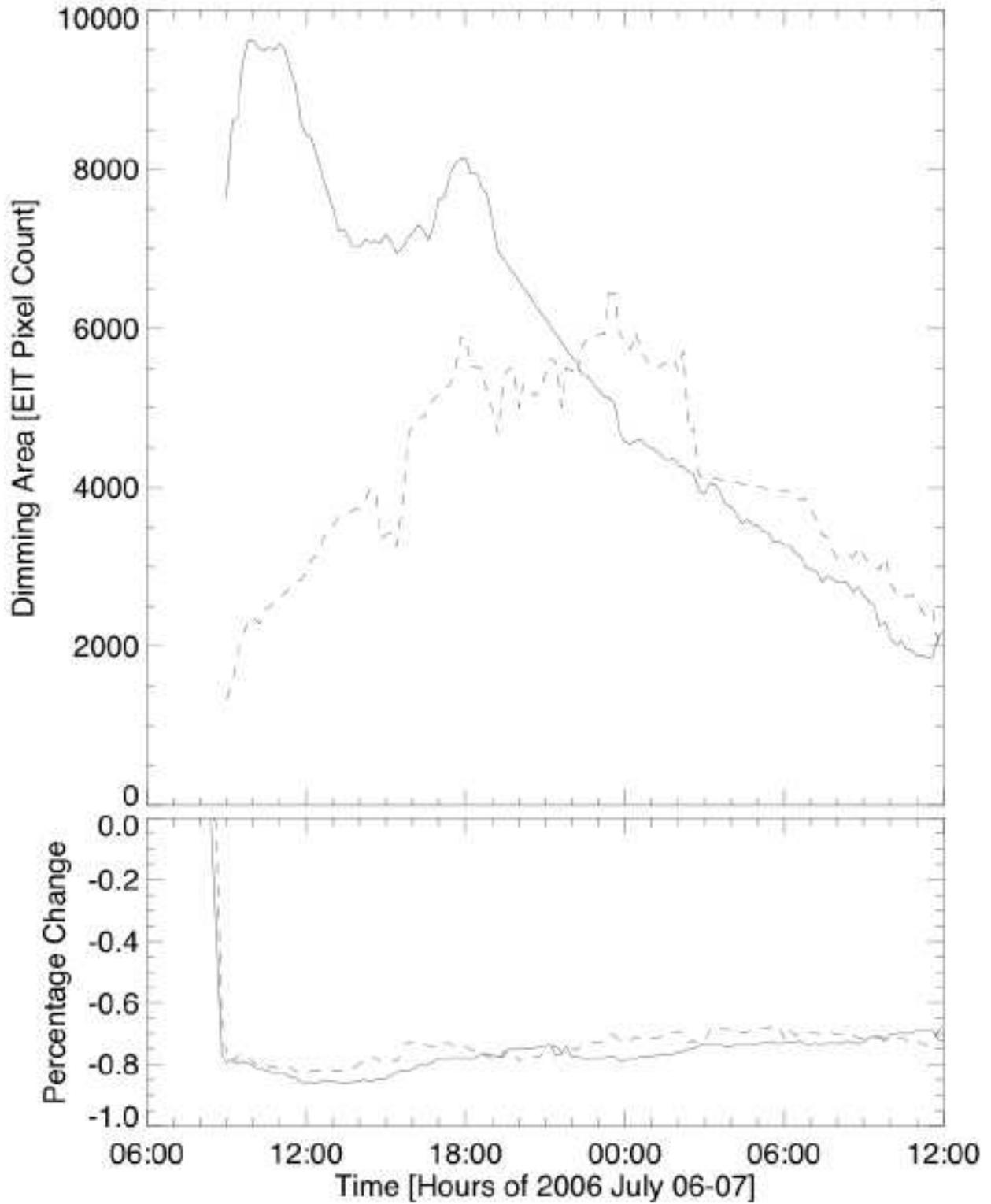}
\caption{Evolution of the transient coronal holes as observed by EIT.  (Top) Time series of the size of lobe 1 (dashed line) and lobe 2 (solid).  (Bottom) Percentage change in intensity from the pre-eruption baseline image (averaged over all pixels in the dimming region). \label{fig3}}
\end{figure}

\begin{figure}
\plotone{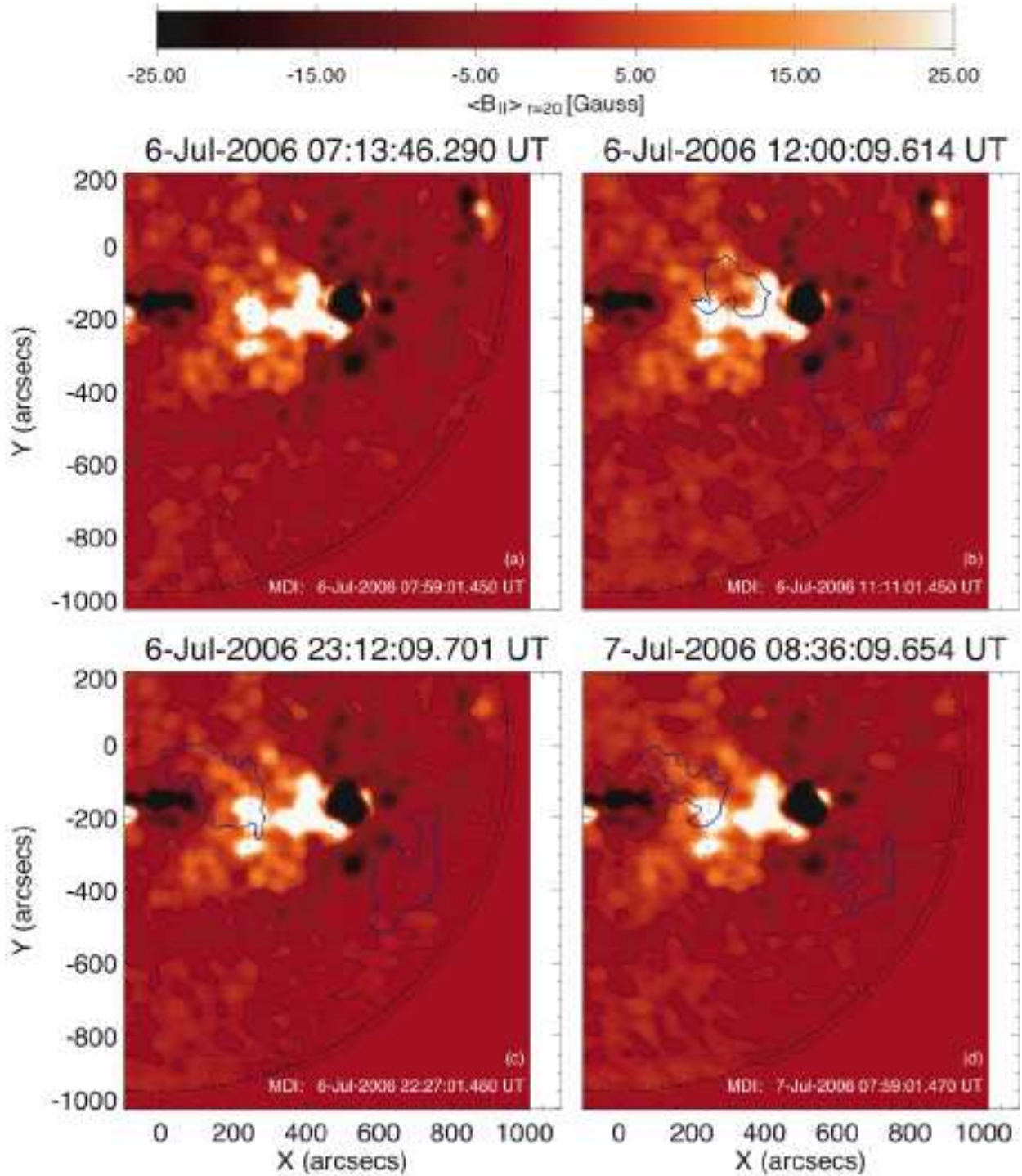}
\caption{Evolution of the SOHO/MDI derived \brr{} over the course of the coronal dimming. The blue contours shown in the panels of the figure represent a 40\% reduction in the EIT intensity while the black contour represents the neutral lines of the \brr{} map (\brr=0). The time of the SOHO/MDI data used to build these panels are shown in the lower right of each panel. [See the electronic edition of the Journal for an mpeg animation of this figure.] \label{fig4}}
\end{figure}

\begin{figure}
\plotone{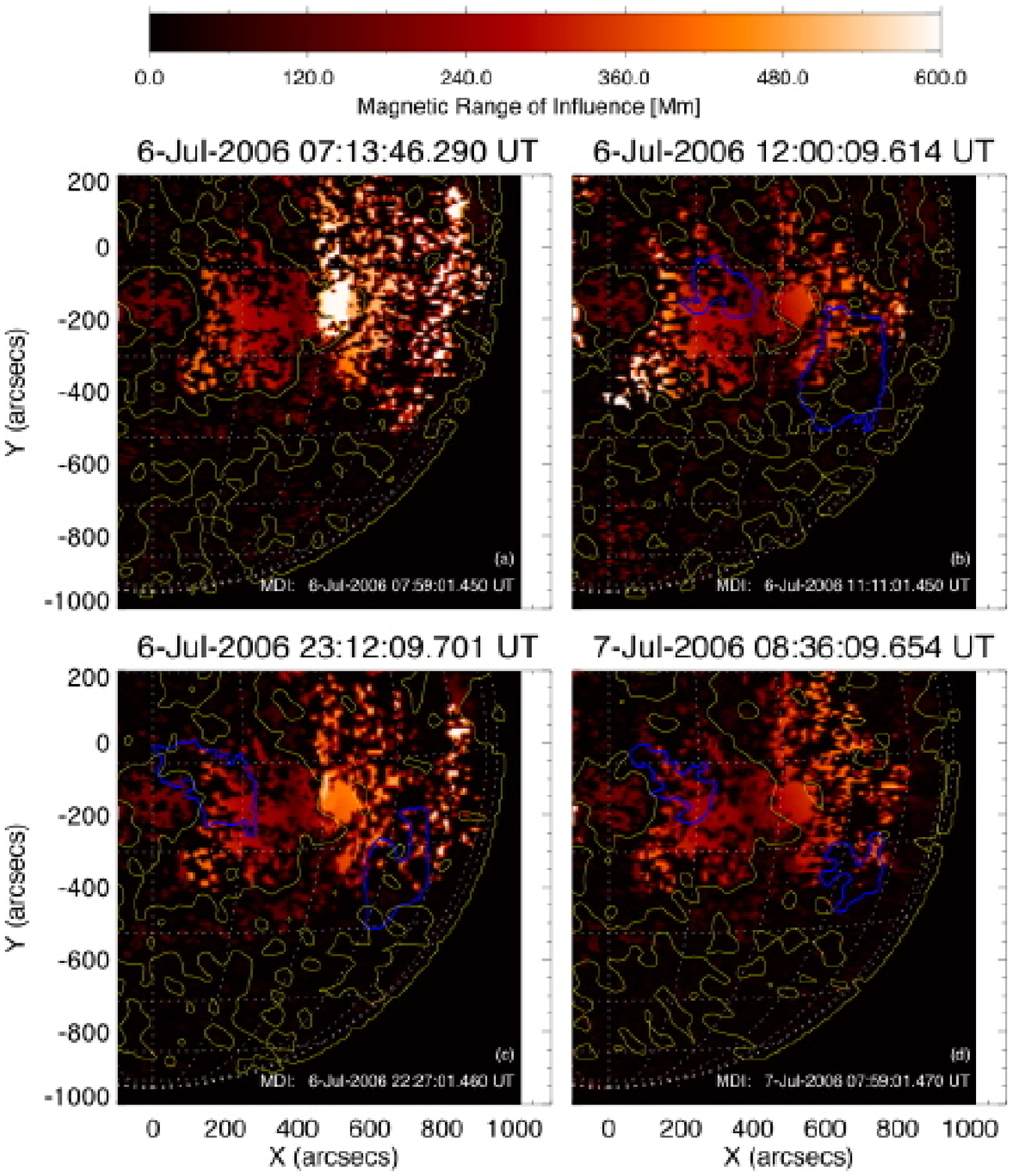}
\caption{Evolution of the SOHO/MDI derived Magnetic Range of Influence (MRoI) over the course of the coronal dimming. The blue contours shown in the panels of the figure represent a 40\% reduction in the EIT intensity while the yellow contour represents the neutral lines of the \brr{} map (\brr=0) in Fig.~\pref{fig4}. The time of the SOHO/MDI data used to build these panels are shown in the lower right of each panel. [See the electronic edition of the Journal for an mpeg animation of this figure.] \label{fig5}}
\end{figure}

The growth and contraction of the dimming regions is shown in Fig.~\pref{fig3}.  The top panel shows the evolution of the area (number of EIT pixels) inside the -40\% contours shown in Fig.~\pref{fig1}. We see that lobe 1 grows steadily until the end of 2006 July 6, whereas lobe 2 starts to contract almost as soon as it is created. The bottom panel shows the time history of the percentage change in intensity. Even though we set the contours in Figures~\pref{fig1}, \pref{fig2}, \pref{fig4} and~\pref{fig5} to be -40\%, the actual average change is closer to -80\%.  Notice that even as the holes fill in (the areas decrease), the intensity of those pixels still inside the -40\% contour remains virtually unchanged.

The spectroscopic analysis of \cite{McIntosh2006a, McIntosh2006b} presents evidence connecting small-scale ejecta driven by magnetoconvection to the occurrence of spicules \cite[e.g.,][]{Beckers1968}. Furthermore, these papers demonstrate how the energy released by the ejecta/spicule is controlled by the closure of the global magnetic field. In the case of the zero-mean field (\brr = 0G), magnetically closed (low MRoI) quiet Sun plasma, the injected spicules contribute to the mass loading and thermal heating of the  because they cannot ``escape'' the confines of the overlying magnetic topology \cite[cf.][]{DePontieu1999}. Conversely, in the unbalanced mean-field (\brr $>$ 5G), magnetically open (high MRoI) coronal hole plasma, the injected spicules have a higher probability of ``escape''. Fueled by the energy in the unbalanced field below, the energy released by the reconnection stays largely kinetic in nature creating the initial plasma acceleration that becomes the fast solar wind. It is hypothesized that the strength of the unbalanced field in the coronal hole sets the scale of the Doppler outflow observed. In subsequent analyses, using long 304\AA{} timeseries observations from EIT, \cite{McIntosh2007a} \cite[and][in preparation]{McIntosh2007b} deduce that large rapid intensity fluctuations in the timeseries result from the same relentless destruction of emerging and advecting magnetic flux and are possibly the transition region emission resulting from dynamic (shock) spicule/fibril formation \cite[e.g.,][]{Hansteen2006, DePontieu2007} or from a different ``species'' of impulsive, purely reconnection-driven, spicules \cite[e.g.,][]{Sterling2000}. \cite{McIntosh2007a} demonstrates that the number, and hence the rate, of these rapid intensity fluctuations increases with the unsigned magnetic flux of the region studied. One can think of the rapidly rising probability of immediately destroying a emerging magnetic flux element (ideally a flux-balanced dipole) in a region where the mean background field is increasing. The enhanced ($\sim$60-70\% increase) ejecta/spicule creation is most evident in the plage surrounding the poles of active regions \cite[e.g., Figs. 3 and 4 of][]{McIntosh2007a} although the correlation extends to coronal holes and quiet Sun \cite[][]{McIntosh2007b}. The consistency of the spatial and spectral patterning present in the various observations studied in \cite{McIntosh2006a, McIntosh2006b, McIntosh2007a, McIntosh2007b} highlights the relentless nature of magnetoconvection and that the process is only quelled in the immediate vicinity of sunspots, never switching off elsewhere on the Sun. As a result we propose that this energy and mass injection process is relevant to the discussion of this paper and governs the apparent removal and refilling of EUV emission that is responsible for the coronal dimming observed.

Invoking the simple energy partitioning controlled by the global magnetic environment \cite[][]{McIntosh2006b} we can explain the post-CME evolution of the coronal dimming. When the top portion of the coronal magnetic topology is stretched out (removed) above the active region by the CME and filament eruption the plasma below ``feels'' an open magnetic field above. As a result the nature of spicular energy delivery to the plasma changes, reflecting the topology change. This change from a thermal to a kinetic energy dominance and opening of the field above ensures that moss begins to disappear.  Similarly, the darkening of the EIT 195\AA{} images indicates that the column depth of the coronal plasma in the region is reducing, i.e., mass is being removed and no longer being replaced.  This implies that the kinetic energy flux now dominates in the portions of the region that are not connected and we can establish the fast speed solar wind streams immediately following CMEs as observed by \cite{Rust1983} \cite[and][]{Skoug2004}. This is exactly what one observes in a coronal hole \cite[][]{McIntosh2006a, McIntosh2006b}.

As the global coronal magnetic field closes from above by reconnection, following the ``disconnection'' of the CME \cite[e.g.,][]{Forbes2000, Lin2000}, the plasma below no longer ``feels'' an open field above and so the spicular ejecta become trapped again and the energy balance returns to favor the thermal component. The dimming lobes begin to fill from beneath stochastically, but with an enhanced rate in regions where unbalanced magnetic flux is available. Indeed, one can see that the re-emergence of moss begins to refill on the same supergranular ``network'' pattern as existed before the eruption (the region [300:450\arcsec, -200:-100\arcsec] shows this clearly) gradually creeping from the neutral line and outward from the active region. It is this pattern of filling that appears to make dimming lobe 1 appear to migrate to the North East. The region to the South West does not show this rapid refilling and migration and we need to know more about the magnetic environment before we can address the apparent migration and stalling of the dimming regions.

We can explain the EIT and TRACE observations of a coronal dimming and complete a self-consistent picture of its post-CME evolution using the simple magnetic diagnostics (\brr, MRoI) introduced by \cite{McIntosh2006a}. In the panels of Figs~\pref{fig4} and~\pref{fig5} (with the associated movies) we show the evolution of the 40\% reduction contour in EIT emission with (co-spatial/de-rotated) maps of \brr{} (Fig.~\pref{fig4}) and MRoI (Fig.~\pref{fig5}) at the same 4 times as Fig.~\pref{fig1}. We should note that the \brr{} and MRoI maps used in this paper are based on spatially averaged values from a line-of-sight magnetograms and should not suffer significantly from the proximity of the active region (or CME) to the west limb. This is true with the possible exception of data within 100\arcsec{} of the data in panel D of both figures that must be de-rotated nearly 24 hours to the time of the dimming appearance.

From Fig. 4, we can see that, as the dimming develops, the contours evolve away from the regions, where \brr $>$ 5G, that form the poles of the active region. In the case of lobe 2 there is very little unbalanced magnetic flux immediately available to rapidly fill the region and so it appears to stall in place not ``moving'' or shrinking appreciably. The situation is very different for the North-Eastern dimming patch where there is a substantial source of unbalanced flux to rapidly fill the area, as discussed above, and it appears to move to the East. Eventually, the apparent migration of this region stops over the region where there is insufficient unbalanced flux to rapidly ``feed'' it.  This hypothesis is supported by the bottom panel of Fig.~\pref{fig3}: the individual pixels fill when exposed to unbalanced flux, but until they are exposed, they remain uniformly dark. The data shown in Fig.~\pref{fig5} explains why this happens: The stalling of both dimming regions occurs when they meet regions of small-scale closed magnetic flux (small MRoI) to the East and West of the active region complex, where we have seen that \brr $\sim$ 0G. These appear to take the form of filament channels (see also Rust 1983). It seems that the MRoI distribution around the active region sets the maximum perimeter of the dimming regions; encircling the maximum amount of magnetic flux that might eventually be seen in a magnetic cloud resulting from the CME \cite[][]{Webb2000, Attrill2006}.

\section{Conclusion}
The asymmetric appearance and evolution of the two dimming patches is entirely dependent on the amount of unbalanced magnetic flux present in the active region following the CME and how that flux is distributed. The factors affecting the evolving dimming patches (different filling rates, different spatial extent, different stalling times) are intimately tied to the relentless action of magnetoconvection and how it is ``driven'' by the flux distribution around the active region. The simple model proposed, while currently lacking exact microphysics, can explain the post-CME evolution of a coronal dimming self-consistently.

Further, the observations and discussion presented above suggest that the use of the term ``transient coronal hole'' to describe a coronal dimming is more than a simple coincidence. For a finite amount of time (from time of CME eruption to the time that the CME tail disconnects from the Sun) the dimming behaves exactly like a coronal hole. This makes the possibility of fast solar wind streams following immediately behind CME associated with the coronal dimming real \cite[][]{Rust1983}. This has interesting implications for the CME modeling community: Is there abundant momentum in the following wind stream to act as a secondary acceleration mechanism of the CME? Unfortunately, from an energy and momentum measuring viewpoint, the eruption studied here was not Earth-directed and so we cannot answer this question in this Paper. However, recent analysis \cite[][]{Qiu2005,Qiu2007} shows a good correlation between MDI flux contained in dimming masks and the initial (SOHO/LASCO measured) velocity of the CME that indicate a possible connection between the CME and post-CME energy release. While we acknowledge that plane-of-sky measurements of CME velocity are not simply connected to the in situ velocity of the ejecta and the following wind stream, the Qiu et~al. results are encouraging, but will need further study. Finally, in situations involving several large geo-effective CMEs \cite[e.g., the 2003 October/November events][]{Skoug2004} we ask if the excessive spicular evaporated cool chromospheric plasma mass acts as progenitor of enhanced Solar Energetic Particle events for following CMEs?

The properties discussed can be tested with existing catalogues of CMEs and coronal dimmings. In fact, they can be used as predictive tools where, in principal, we will be able to estimate the maximum amount of magnetic flux and energy available around a particular active region. One step of this process has already been performed for one simple event \cite[][]{Webb2000, Attrill2006} and indirectly for several others \cite[][]{Qiu2007}. We hope that a better understanding of the physical process behind the appearance and post-CME evolution of coronal dimmings will enhance the modeling of CME propagation and the general prediction of space weather conditions.

In a companion paper \cite[][in preparation]{McIntoshLeamon2007} we explore the possibility that stochastic magnetoconvection-driven flux emergence and cancellation of the tethering magnetic flux in the weakest flux portions of the active region complex can explain the initiation of the filament eruption and CME itself, not just the post-event evolution.


\acknowledgements
We would like to thank Bart De Pontieu for helpful discussions on the occurrence of TRACE moss and the anonymous referee for helpful comments on the manuscript. The work presented was supported by the National Aeronautics and Space Administration under grants to the author issued from the Solar Data Analysis Center (NNG05GQ70G), Sun-Earth Connection Guest Investigator Program (NNG05GM75G), the Solar and Heliospheric Physics Program (NNG06GC89G). The National Center for Atmospheric Research is sponsored by the National Science Foundation.

\clearpage

\end{document}